\documentclass[showpacs,showkeys,preprintnumbers,amsmath,amssymb,aps]{revtex4}

\usepackage{graphicx}
\usepackage{bm}

\newcommand {\calleq}[1]{(\ref{eq:#1})}
\newcommand {\lab}[1]{\label{eq:#1}}

\begin{document}

\preprint{PREPRINT}

\title{
The Fermi-Pasta-Ulam problem: 
\\
periodic orbits, normal forms
and resonance overlap criteria}

\author{S. Flach$^1$ and A. Ponno$^2$}
 

\affiliation{
$^1$ Max-Planck-Institut f\"ur Physik komplexer Systeme
\\
N\"othnitzer Str. 38, 01187 Dresden, Germany
\\
$^2$ Universit\`a degli Studi di Padova\\
Dipartimento di Matematica Pura ed Applicata
\\
Via Trieste, 63 - 35121 Padova, Italy}

\date{\today}

\begin{abstract}
Fermi, Pasta and Ulam observed, 
that the excitation of a low frequency normal mode 
in a nonlinear acoustic chain leads to
localization in normal mode space on large time scales.
Fast equipartition (and thus complete delocalization) in the Fermi-Pasta-Ulam chain 
is restored if relevant intensive control parameters exceed certain
threshold values. 
We compare recent results on periodic orbits (in the localization regime)
and resonant normal forms (in a weak delocalization regime), and
relate them to various resonance overlap criteria.
We show that the approaches quantitatively agree
in their estimate of the localization-delocalization threshold.
A key ingredient for this transition are resonances of
overtones. 
\end{abstract}


\maketitle

\section{INTRODUCTION}
Fermi, Pasta and Ulam (FPU) 
considered the dynamics of a classical nonlinear acoustic chain 
of particles with coordinates $x_n$, canonically conjugated momenta $p_n$ 
and the Hamiltonian
\begin{equation}
 H=\sum_{n=0}^N \left[ \frac{p_n^2}{2}+\frac{(x_{n+1}-x_n)^2}{2}+
 \alpha \frac{(x_{n+1}-x_n)^3}{3}
 +\beta \frac{(x_{n+1}-x_n)^4}{4}  \right] \;,
\lab{H1}
\end{equation}
where $\alpha$ and $\beta$ are the nonlinearity parameters.   
Fixed boundary conditions  $x_0=x_{N+1}=p_{0}=p_{N+1}=0$ are assumed. 
$N$ is the number of degrees of freedom of the system. 
FPU considered
the case with $\alpha>0$ and $\beta=0$, and the one with $\alpha=0$ and $\beta>0$.
These two models \calleq{H1} are known as the $\alpha$--model and $\beta$--model,
respectively.

FPU performed a numerical experiment \cite{FPU} with the aim to observe an approach to 
an \emph{expected} energy equipartition among the Fourier modes of system \calleq{H1}, when the 
longest wavelength mode is initially excited. Instead,
they discovered that the energy, initially given to the 
longest wavelength mode, is partially transferred just to a few modes of shorter 
wavelength that are excited one after the other in order of decreasing wavelength. Such an 
energy cascade was observed to be effective, in practice, up to a certain critical mode, 
giving thus rise to a state of the system quite different
from the equipartitioned one and apparently persisting up to their largest integration time, 
in the range of parameters (energy, number of degrees of freedom, nonlinearity 
parameters) they chose. Such an observed lack of complete energy equipartition among the 
modes of the system up to unexpectedly long times posed a problem which, after them, was 
named the FPU problem or paradox. We stress here, that
the observation of an energy distribution in normal mode space staying \emph{localized}
for just the times on which nonequipartition is registered, is perhaps \emph{the}
main issue of the paradox. Indeed long-studied questions of estimating recurrence times
are already based on the fact that the energy \emph{stays} localized in normal mode space
for long times \cite{Ford92}.  
Galgani and Scotti \cite{GS} observed numerically in 1972, that the localization
profile is exponential.

The basic mechanism of the energy cascade in the FPU problem was 
identified by Ford (1961)
\cite{F}. The long wavelength modes 
are characterized by a dispersion relation of the type 
$\omega_q\simeq  \pi q/N$, 
where $q=1,\dots,N$ is the mode index, so that
\begin{equation}
 \omega_1\simeq\omega_2/2\simeq\omega_3/3\dots\ \ .
\lab{acres}
\end{equation}
Due to such an almost complete acoustic resonance condition a resonant excitation of consecutive 
modes of decreasing wavelength, or increasing frequency, is expected. Notice that the approximate 
equalities \calleq{acres} hold with an error
increasing with $q$, so that one can expect the resonant pumping mechanism to become uneffective at a certain
critical $q_c$.

In 
1966 Izrailev and Chirikov (IC) \cite{IC} realized that increasing the energy of the initially excited mode will lead to
a crossover from the localization (nonequipartition) regime to a
delocalization (equipartition) one, 
where a fast transfer of energy from long to short 
wavelength modes will lead to energy equipartition, as originally expected by FPU. 
IC used a nonlinear resonance overlap criterion to estimate the crossover.
Notably they considered resonances different from \calleq{acres}.

Both the work of IC and the idea of Ford have been re-considered by Shepelyansky \cite{Sh}
in 1997, who applied the IC idea by using the Ford resonances \calleq{acres}.
He obtained a resonant normal form for the system \calleq{H1}, 
from which he arrived at an estimate 
of the critical mode index $q_c$ at which the cascade effectively slows down. Numerical runs
yielded that the numerical 
distribution of the harmonic energies $E_q$ of the modes (i.e. $E_q$ vs. $q$) displays an 
exponentially decreasing tail of the form $E_q\sim\exp(-q/q_c)$. 
Since Shepelyansky used resonances which are different from the ones originally
used by Izrailev and Chirikov,
the crossover estimate from \cite{Sh} does not coincide with the
IC result.

Two more recent lines of research are the reason for this report.
First, a normal mode at the linear limit $\alpha=\beta=0$, when excited to a given energy, represents
a periodic orbit in the phase space of the system \calleq{H1}.
That periodic orbit persists for nonzero $\alpha,\beta$ at the same energy
and is coined $q$-breather.
Existence proofs, perturbation theory and high precision numerical calculations
\cite{FIKa,FIKb,IKMF,KFIM,PF,FKMI} allow to draw a consistent picture of these periodic orbits.
For weak nonlinearity $q$-breathers are exponentially localized in normal mode space.
Analytical estimates allow to conclude about the existence of a delocalization threshold
in the control parameters,
above which the $q$-breather delocalizes. Small deviations from these orbits stay
close to them for sufficiently long times. Thus some features of the 
trajectory first computed by FPU are quantitatively captured by the properties of a
$q$-breather and by the linearized phase space flow around it. 
The scaling properties of
$q$-breathers appear to be close to the results of Shepelyansky.

Second, a careful reconsideration of the resonant normal form 
\cite{P1,P2,PB1,PB2,BP1,BP2} allowed to get analytic estimates of the modal 
energy spectrum of the system and of the time scales involved in the problem. The 
direct relevance of the work of Zabusky and Kruskal \cite{ZK} on the KdV equation to the FPU 
problem has been almost completely clarified. Again some of the scaling properties of the
normal form are similar to those of Shepelyansky, and thus related to the 
scaling of $q$-breathers. That is remarkable, since $q$-breathers and normal forms
seem to operate in conceptually different regimes, namely the localization and weak delocalization
regime, respectively.

Here we intend to compare all these results with each other, in order to
establish the similarities and differences. The emerging picture is,
that the relaxation properties of a normal mode, launched into a nonlinear acoustic
lattice, are governed by intensive quantities (i.e. energy density, frequency, etc).
Below certain parameter thresholds the relaxation is slow, while above them it is fast,
and that circumstance may survive in the thermodynamic limit.

\section{Transformation to normal modes}

The usual canonical change of variables $(x,p)\mapsto(Q,P)$ that transforms Hamiltonian \calleq{H1} 
into the sum of a Hamiltonian of noninteracting mode oscillators plus a perturbation responsible 
for mode coupling is given by
\begin{eqnarray}
 && Q_q=\sqrt{\frac{2}{N+1}}\ \sum_{n=1}^Nx_n\sin\left(\frac{\pi qn}{N+1}\right)\;, \\
 && P_q=\sqrt{\frac{2}{N+1}}\ \sum_{n=1}^Np_n\sin\left(\frac{\pi qn}{N+1}\right)\ \ .
\end{eqnarray}
for $q=1,\dots,N$. In terms of the canonical normal mode variables $(Q,P)$ 
the Hamiltonian
\calleq{H1} reads
\begin{eqnarray}
 H(Q,P) & = &
 \sum_{q=1}^N\frac{P_q^2+\omega_k^2Q_q^2}{2}+\frac{\alpha}{3\sqrt{2N+2}}
 \sum_{j,q,l=1}^NS_3(j,q,l)\omega_j\omega_q\omega_lQ_jQ_qQ_l +\nonumber \\
 && + \frac{\beta}{8(N+1)}\sum_{j,q,l,m=1}^NS_4(j,q,l,m)\omega_j\omega_q\omega_l\omega_mQ_jQ_qQ_lQ_m\ \ ,
\lab{H2}
\end{eqnarray}
where 
\begin{equation}
 \omega_q=2\sin\left(\frac{\pi q}{2N+2}\right)
\lab{disprel}
\end{equation}
is the dispersion relation (frequency vs. mode index) of the normal mode of the noninteracting system,
whereas the two coefficients $S_3$ and $S_4$ appearing in \calleq{H2} rule the 
actual mode coupling in the system, and are given by
\begin{equation}
 S_3(j,q,l) = \delta_{j+q,l}+\delta_{j+l,q}+\delta_{q+l,j}-\delta_{j+q+l,2N+2} \;,
 \lab{s3}
\end{equation}
\begin{eqnarray}
 S_4(j,q,l,m) & = & \delta_{j+q+l,m}+\delta_{q+l+m,j}+\delta_{l+m+j,q}+\delta_{m+j+q,l} + \nonumber\\
 && \delta_{j+q,l+m}+\delta_{j+l,q+m}+\delta_{j+m,q+l}-\delta_{j+q+l+m,2N+2} +\nonumber\\
 && -\delta_{j+q+l,m+2N+2}-\delta_{q+l+m,j+2N+2}-\delta_{l+m+j,q+2N+2}-\delta_{m+j+q,l+2N+2} \lab{s4}\ \ ,
\end{eqnarray}
$\delta_{n,m}$ denoting the usual Kronecker symbol. 
The coefficients $S_3$ and $S_4$ are invariant under permutation of any pair of indices. 
Notice that the dispersion relation \calleq{disprel}, when $q/(N+1)$ is small, admits the expansion
\begin{equation}
 \omega_q=\frac{\pi q}{N+1}-\frac{1}{24}\left(\frac{\pi q}{N+1}\right)^3+O((q/N)^5)\ \ ,
\lab{omexp}
\end{equation}
so that, in a very first approximation, low frequency modes 
(small $q$) display a quasi-linear dependence of the frequency on the
mode index and their frequencies approximately satisfy the relation \calleq{acres}.
The equations of motion associated to the Hamiltonian \calleq{H2} read
\begin{eqnarray}
 \ddot{Q}_q+\omega_q^2Q_q & = & -\frac{\alpha\omega_q}{\sqrt{2N+2}}
 \sum_{j,l=1}^NS_3(q,j,l)\omega_j\omega_lQ_jQ_l+ \nonumber\\
 && -\frac{\beta\omega_q}{2N+2}\sum_{j,l,m=1}^NS_4(q,j,l,m)\omega_j\omega_l\omega_mQ_jQ_lQ_m\ \ .
\end{eqnarray}
They are used for constructing
$q$-breathers.
For what concerns the normal form approach, a further canonical transformation to complex variables
turns out to be useful. More precisely, one defines the change of variables $(Q,P)\mapsto(z,z^*)$ by
\begin{equation}
 z_q=\frac{\omega_qQ_q+iP_q}{\sqrt{2\omega_q}}
\lab{zk}
\end{equation} 
and its complex conjugate $z_q^*$ for any $q=1,\dots,N$. Notice that 
\begin{equation}
E_q \equiv \omega_q|z_q|^2=(P_q^2+\omega_q^2Q_q^2)/2 \lab{harmen}
\end{equation}
is the harmonic energy of mode $q$, so that $|z_q|^2$ is the action variable associated to the same mode.
Regarding $z_q$ as a canonical coordinate, it turns out that its conjugate momentum is given by $iz_q^*$, since
\[
 \{z_r,z_s^*\}=-i\delta_{r,s}\ \ ,
\]
where $\{...,...\}$ denotes the usual canonical Poisson bracket.
Thus the Hamiltonian \calleq{H2} becomes
\begin{eqnarray}
 H=\sum_{q=1}^N\omega_q|z_q|^2+\frac{\alpha}{12\sqrt{N+1}}\ \sum_{j,q,l=1}^NS_3(j,q,l)
 \sqrt{\omega_j\omega_q\omega_l}(z_j+z_j^*)(z_q+z_q^*)(z_l+z_l^*)+\nonumber\\
 +\frac{\beta}{32(N+1)}\ \sum_{j,q,l,m=1}^NS_4(j,q,l,m)\sqrt{\omega_j\omega_q\omega_l\omega_m}
 (z_j+z_j^*)(z_q+z_q^*)(z_l+z_l^*)(z_m+z_m^*)\ \ ,
\lab{H3}
\end{eqnarray}
with equations of motion given by $i\dot{z}_q=\partial H/\partial z_q^*$.

\section{The FPU problem}

\begin{figure}
{
\resizebox*{0.5\columnwidth}{!}{\includegraphics{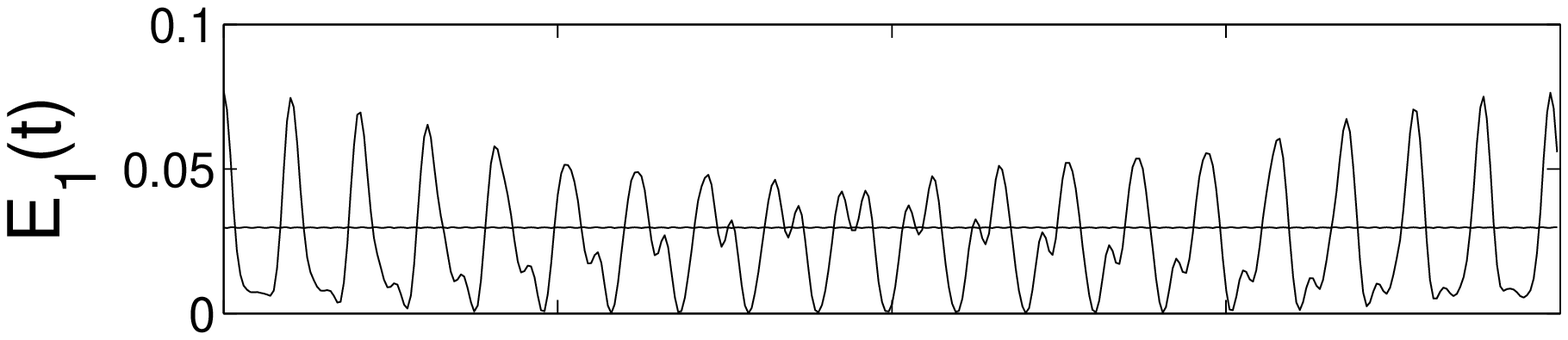}}
\resizebox*{0.5\columnwidth}{!}{\includegraphics{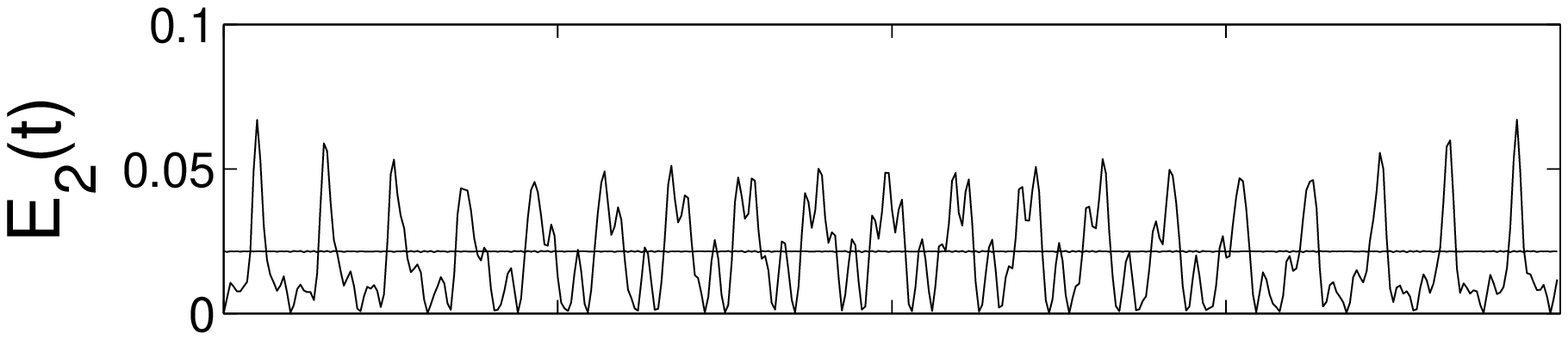}}
\resizebox*{0.5\columnwidth}{!}{\includegraphics{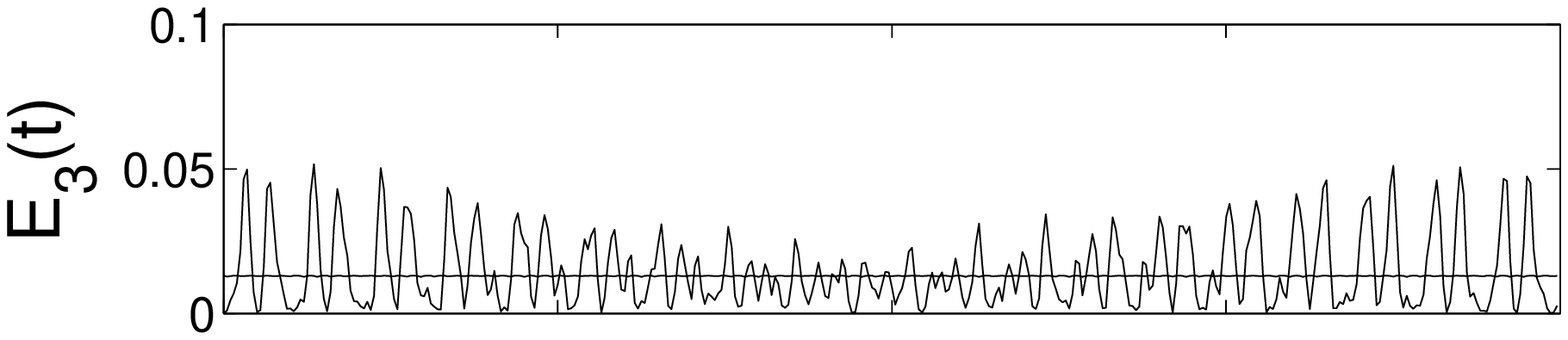}}
\resizebox*{0.5\columnwidth}{!}{\includegraphics{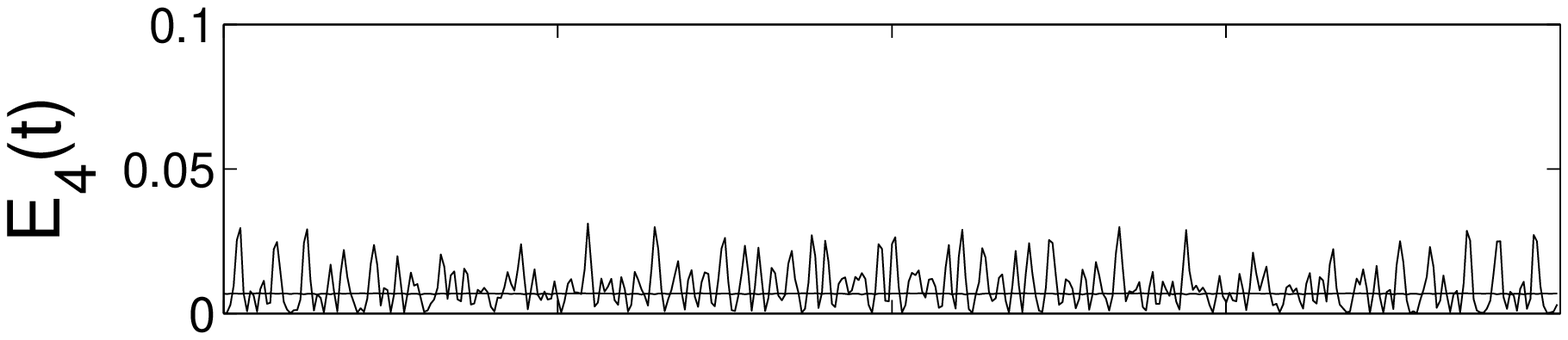}}
\resizebox*{0.5\columnwidth}{!}{\includegraphics{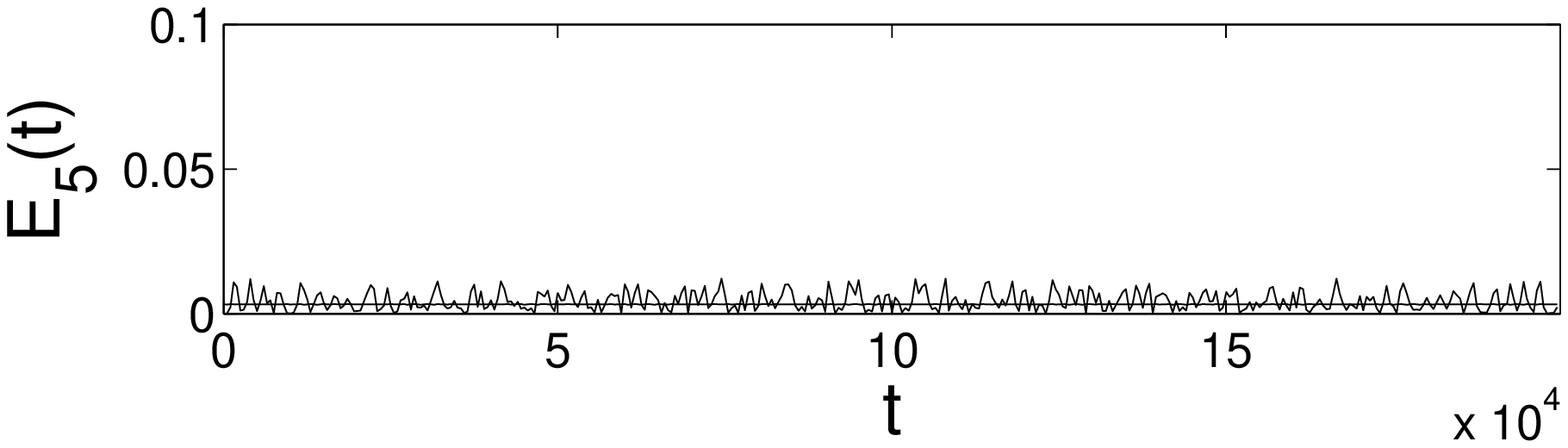}}}
\caption{
Normal mode energy evolution for $q=1,2,3,4,5$ for (i) an FPU trajectory
(oscillating curves) and (ii) an exact $q$-breather periodic orbit (horizontal
lines). Parameters are $\alpha=0.25$, $E=0.077$, $N=32$ and $q_0=1$.
From \cite{FIKb} }
\label{Fig1}
\end{figure}

Fermi, Pasta and Ulam launched a trajectory by initially exciting just one
normal mode with seed mode number $q_0$, and monitored the time evolution of the harmonic mode energies
\calleq{harmen}. A typical outcome of their experiment is shown in Fig.\ref{Fig1}.
Instead of spreading among all normal modes, the energy stays localized
on a few long wavelength modes. Recurrence events are observed on times
much shorter than the expected Poincar\'e recurrence times. At the same
time the localization of energy in normal mode space happens for times
which are orders of magnitude larger than the recurrence times.

\begin{figure}
{
\resizebox*{0.9\columnwidth}{!}{\includegraphics{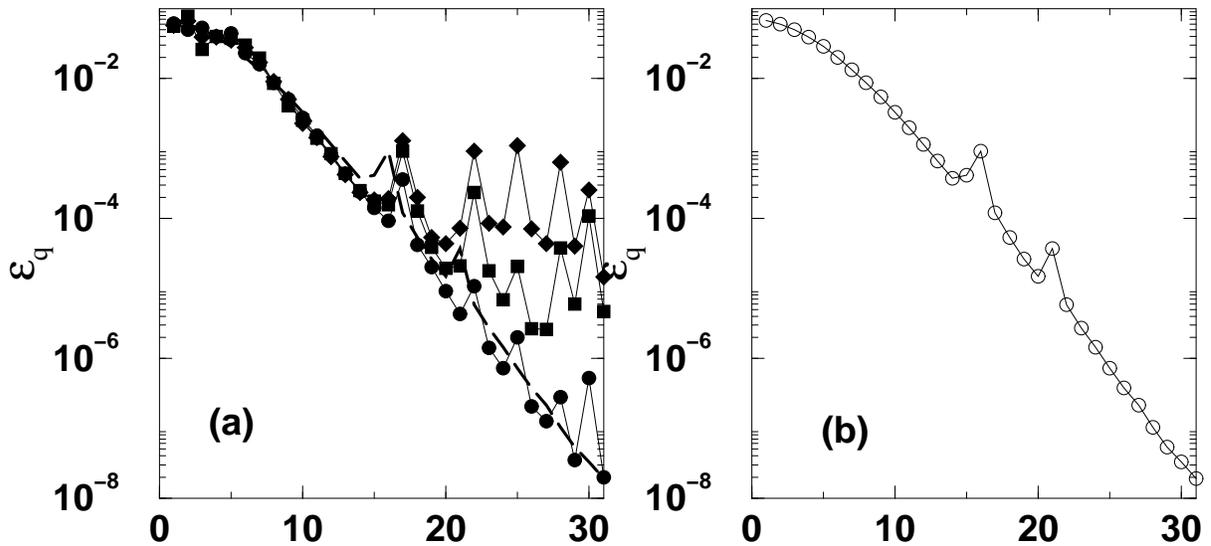}}}
\caption{
(a) Distribution of normal mode energy densities for an FPU trajectory 
and parameters $q_0=1$, $N=31$, $\alpha=0.33$, $E=0.32$.
Circles - $t=10^4$, squares - $t=10^5$, rhombs - $t=10^6$.
Dashed line - distribution for the $q$-breather periodic orbit from (b)
for comparison. (b) Distribution of normal mode energy densities
for a $q$-breather periodic orbit and the parameters as in (a).
From \cite{PF} }
\label{Fig2}
\end{figure}

In Fig.\ref{Fig2} we show distributions of normal mode energy densities
$\epsilon_q = E_q/N$ for larger initial energies, and at different
times $t=10^4,10^5,10^6$. To eliminate fluctuations,
they are averaged over a preceding time window of $10^4$.
We clearly observe exponential localization of the distribution, 
and note that such a distribution is reached on a quite fast time scale 
$\tau_1 \ll 10^4$. We also observe a slow resonant excitation of tail modes,
which ultimately drags the system to equipartition, but on a 
much larger second time scale $\tau_2 \gg \tau_1$ \cite{BGP}.
The origin of these tail resonances is explained in \cite{PF}.

Further increasing e.g. the seed mode energy, or the nonlinearity, may lead to a threshold
where fast delocalization and equipartition take place, i.e. where
$\tau_1 \approx \tau_2$ \cite{BGP}. Below these thresholds, the two time scales
are strongly separated, and the evolution on times $\tau_1 \ll t \ll \tau_2$ corresponds
to the regime observed by FPU. Such a double time-scale scenario is akin to that
proposed first in \cite{Fetal}. 
From this perspective, the FPU problem in a nutshell may be phrased as:
{\sl Why are there (at least) two different time scales, how to estimate the
localization length and what is its relation to the time scales, is the effect robust
in the limit of infinitely large system sizes, as well as for higher spatial dimensions?}

\section{Continuation of normal modes: $q$-breathers}

Let us assume, that the FPU trajectory is close to a periodic orbit (PO)
for times $t \ll \tau_2$. Then the FPU trajectory evolution will be almost regular, as
observed in the simulations. It can be expected, that the PO will correspond
to a slightly deformed normal mode solution of the linear case \cite{ford63}. 

\subsection{General properties}

A rigorous construction scheme of such POs (including an existence proof)
was given in 
\cite{FIKa,FIKb}, which starts from the simple linear limit $\alpha=\beta=0$.
Essentially one chooses a seed mode number $q_0$ and an energy $E$.
This choice uniquely defines a periodic orbit in the phase space of \calleq{H1}
for $\alpha=\beta=0$. Moreover, for any finite size $N$ the nonresonance condition
$n\omega_{q_0} \neq \omega_{q \neq q_0}$ holds for any integer $n$ \cite{FIKb}.
Therefore straight application of a persistence theorem of periodic orbits by Lyapunov \cite{Ly}
guarantees the continuation of that periodic orbit for nonzero $\alpha$ and $\beta$.
Notice that for uniqueness one has to fix one parameter, e.g. the total energy $E$.

The continued periodic orbits are exponentially localized in normal mode space
and follow
the corresponding FPU trajectories impressively closely (cf.
Fig.\ref{Fig1} and Fig.\ref{Fig2}). They are 
coined $q$-breathers \cite{FIKa,FIKb,IKMF,KFIM,PF,FKMI}. The localization length tends to increase
with increasing nonlinearity, or increasing energy, or decreasing seed mode number.
Numerical high-precision schemes for computing these orbits are given in \cite{FIKa,FIKb}.

Notably $q$-breathers may stay stable, or also turn unstable, keeping their localization \cite{FIKa,FIKb}.
Such instabilities, which are related to the emergence of weak chaos in the core of the corresponding
FPU trajectory \cite{LLL},
do not affect the time scale $\tau_2$ to a significant extend. That is reasonable, since
an instability marks a resonance, but the fact that the $q$-breather is well localized tells then,
that the resonance can not be between core and tail (which would lead to delocalization in the first
place). Thus the only possibility is a resonance between two modes in the core, which changes the
core dynamics, but leaves the localization properties almost unchanged.

The concept of $q$-breathers has been successfully applied to higher dimensional systems
as well, without much change in their properties \cite{IKMF}. 
We think that most of the results to be discussed below, though
strictly speaking being correct only for one spatial dimension, will hold in higher
dimensions as well.

\subsection{Scaling to large sizes}

Application of Poincar\'e-Lindstedt perturbation theory to $q$-breathers \cite{FIKa,FIKb,IKMF}
suggests that only intensive quantities matter, namely the energy density $\epsilon$ and the wave number $k$
defined by
\begin{equation} 
\epsilon = E / N\;,\; k = \pi q /(N+1)
\label{intensivevar}
\end{equation}
matter. It follows that $q$-breathers may persist in infinitely large systems.
As was shown in \cite{KFIM}, it is enough to first obtain a $q$-breather for a chain with
some finite $N$. Let us then consider a new chain with size $\tilde{N} +1 = r(N+1)$, with
$r=2,3,4,...$. It follows, that 
\begin{equation}
\label{scaling1}
\tilde{Q}_{\tilde{q}}(t)=
\left\{
\begin{array}{ll}
\sqrt{r} Q_q(t), & \tilde{q}=r q
\\
0, & \tilde{q} \neq r q
\end{array}\;,\right.
\end{equation}
is a solution of the scaled chain. 
Fixing $r$ and repeating that procedure, or simply considering the limit $ r \rightarrow \infty$,
we will obtain solutions for macroscopic systems. These scaling procedures are easily 
generalized to two and three spatial dimensions, as well to other boundary conditions \cite{KFIM}.

\subsection{Localization properties: $\beta$ FPU}

\begin{figure}
{
\resizebox*{0.5\columnwidth}{!}{\includegraphics{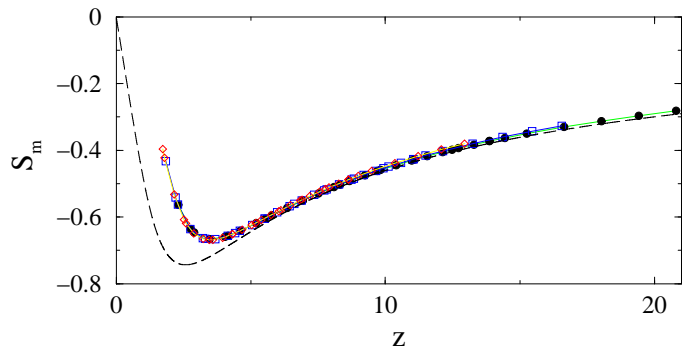}}}
\caption{
$S_m(z)$ (dashed line). Symbols and guiding line: data from $q$-breather calculations
for systems with $\beta=1$, $\epsilon=6.08 \cdot 10^{-4},9.6\cdot 10^{-4},1.57\cdot 10^{-3}$
and $N=149,359$.
From \cite{KFIM} }
\label{Fig3}
\end{figure}

The perturbation theory results \cite{FIKa,FIKb} for the FPU $\beta$-model yield the following form for the modal energy spectrum 
of a $q$-breather with seed wave number $k_0$ \cite{KFIM}:
\begin{equation}
\label{localizationbeta}
\ln \epsilon_k = 
\left(\frac{k}{k_0}-1 \right) \ln \sqrt{\lambda} + \ln \epsilon_{k_0}
\;,\; 
\sqrt{\lambda}=\frac{3\beta}{8} \frac{\epsilon_{k_0}}{k_0^2}\;.
\end{equation}
Here $\epsilon_{k_0}$ is the seed mode (harmonic) specific energy.
Localization holds for $\lambda < 1$, which is a function of intensive quantities only.
It takes an exponential law 
\begin{equation}
\epsilon_k \sim {\rm e}^{-k/\xi}
\label{localizationlength}
\end{equation}
where $\xi$ is the localization length (in $k$-space).
We consider $q$-breathers at
fixed average energy density $\epsilon$. Using (\ref{localizationbeta}) 
it follows $\epsilon_{k_0} = (1-\lambda)\epsilon$.
Together with the definition of $\lambda$ in (\ref{localizationbeta}) we find,
that the inverse of the localization
length $\xi$ in $k$-space is given by:
\begin{equation}
\xi^{-1}= \frac{-1}{k_0} \ln \sqrt{\lambda}\;,\;
\sqrt{\lambda} = \frac{z^2}{2} (\sqrt{1+4z^{-4}}-1)\;,\;z=\frac{k_0}{\nu}\;,\;
\nu^2=\frac{3\beta}{8}\epsilon\;,
\label{slope}
\end{equation}
where $z$ is a scaled wavenumber and $\nu$ the effective nonlinearity parameter.
It follows, that $\xi^{-1}=-S_m/\nu$ where the master function $S_m=z^{-1} \ln \sqrt{\lambda} $ depends
only on the scaled wavenumber $z$.
$S_m$
vanishes for $z \rightarrow 0 $ and has its largest absolute value
$max(|S_m|) \approx 0.7432$  at
$z_{min}\approx 2.577$.
For a fixed effective nonlinearity parameter $\nu$
the $q$-breather with $k_{min}=\nu z_{min}$ shows the strongest localization,
while larger and smaller $k_0$ tend to weaken the localization.
Especially for $k_0 \rightarrow 0$ the $q$-breather delocalizes completely.
With increasing $\nu$,
the localization length of the $q$-breather for $k_0=k_{min}$
increases. For $k_0 \gg \nu$ it follows $S_m \approx - 2/z \ln (z)$
and for $k_0 \ll \nu$ we find $S_m \approx -z$ (see Fig.\ref{Fig3}).
All predictions are verified numerically. It follows, that
in the regime of strong localization $\epsilon \approx \epsilon_{k_0}$
the exponential decay in units of $k_0$ depends on the single
parameter $3\beta\epsilon_{k_0}/(8k_0^2)$, implying corresponding
scaling laws.  
In particular, delocalization is expected to be reached when
\begin{equation}
\frac{3\beta}{8} \frac{\epsilon}{k_0^2}=1\;.
\label{betadelocstrong}
\end{equation}
However, for long wavelength it follows 
\begin{equation}
\xi = \nu^2/k_0\;,\;k_0 \ll \nu
\;.
\label{delocalization}
\end{equation}
Since $k$-space is finite, localization is only meaningful if (at most) $\xi < \pi$.
Consequently $q$-breathers delocalize at a threshold given by
\begin{equation}
\frac{3\beta \epsilon}{8 k_0} \approx \pi
\;.
\label{thresholdbeta}
\end{equation}
Clearly the scaling law is now different from the one for strongly localized
$q$-breathers (\ref{betadelocstrong}). 

The reason for the weaker localization of
$q$-breathers when $k_0 \gg \nu$ is the increasing distance $3k_0$  
between modes excited in consecutive orders of perturbation theory.
The delocalization for $k_0 \rightarrow 0$ however is
due to an approaching of resonances $(2n+1)\omega_{k_0} \rightarrow \omega_{(2n+1)k_0}$ for some
integer $n$, which are the Ford resonances \calleq{acres}.

Most importantly, for each value of $\beta$ and $\epsilon$ we can identify a critical
seed wave number 
\begin{equation}
\label{betakc1}
{k}_{c1} = \frac{3 \beta \epsilon}{8 \pi}\;.
\end{equation}
For $k_0 > {k}_{c1}$
$q$-breathers are localized, while for $k_0 < {k}_{c1}$ they delocalize. This delocalization
threshold, corresponding to condition (\ref{thresholdbeta}), can be expected to translate into a delocalization threshold
of a corresponding FPU trajectory. If instead we use the scaling law (\ref{betadelocstrong})
for strongly localized $q$-breathers,
we arrive at a critical seed wave number 
\begin{equation}
\lab{betakc2}
{k}_{c2} = \sqrt{\frac{3 \beta \epsilon}{8}}\;.
\end{equation}

\subsection{Localization properties: $\alpha$ FPU}

In a similar study \cite{FKMI} the $\alpha$-FPU case yields analogous results.
The modal energy spectrum of a $q$-breather with seed wave nuber $k_0$ is given by
\begin{equation}
\label{localizationalpha}
\ln \epsilon_k = \left(\frac{k}{k_0}-1 \right) \ln\gamma + 
+2 \ln \left( \frac{k}{k_0}\right) \ln \epsilon_{k_0}, \quad
\gamma=\frac{\alpha^2 \epsilon_{k_0}}{k_0^4}.
\end{equation}
Note that the exponential decay here is dressed by a weaker but nonzero
power law, which may be important
when it comes to quantitative comparisons.
Using (\ref{localizationalpha}) it follows
$\epsilon_{k_0} = \frac{(1-\gamma)^3}{1+\gamma}\epsilon$
and thus 
\begin{equation}
\label{alphagamma}
\gamma=z^{-4} \frac{(1-\gamma)^3}{1+\gamma}\;,\;z=\frac{k_0}{\phi}\;,\;
\phi=\sqrt{\alpha}\; \epsilon^{1/4}\;.
\end{equation}
Here we again introduced a scaled wave number $z$ and the effective nonlinearity parameter
$\phi$. The inverse localization length $\xi$ of a $q$-breather is then defined
by a master slope $S_m$ via
\begin{equation}
\label{alphaslope}
\xi^{-1} = -\frac{S_m}{\phi}\;,\;S_m = \frac{1}{z} \ln \gamma\;.
\end{equation}
The dependence of $S_m(z)$ is qualitatively similar to the $\beta$-FPU case. It has a minimum
at $z_{min} \approx 2.39$ and a value $S_m(z_{min}) \approx -1.5$.
At variance with the $\beta$-FPU case the dependence for small $z \ll 1$
is $S_m \sim -z^{1/3}$.

All predictions are again verified numerically. It follows, that
in the regime of strong localization $\epsilon \approx \epsilon_{k_0}$
the exponential decay in units of $k_0$ depends on the single
parameter $\alpha^2 \epsilon_{k_0}/k_0^4$, implying corresponding
scaling laws.  
In particular, delocalization is expected to be reached when
\begin{equation}
\frac{\alpha^2 \epsilon_{k_0}}{k_0^4}=1\;.
\label{alphadelocstrong}
\end{equation}
However, for long wavelengths it follows 
\begin{equation}
\xi = \phi^{4/3}/(2k_0)^{1/3}\;,\;k_0 \ll \phi
\;.
\label{delocalizationalpha}
\end{equation}
Since $k$-space is finite, localization is only meaningful if (at most) $\xi < \pi$.
Consequently $q$-breathers delocalize at a threshold given by
\begin{equation}
\frac{\alpha^2 \epsilon}{2 k_0} = \pi^3
\;.
\label{thresholdalpha}
\end{equation}
Again the scaling law is now different from the one for strongly localized
$q$-breathers (\ref{alphadelocstrong}). 

The reason for the weaker localization of
$q$-breathers when $k_0 \gg \phi$ is the increasing distance $2k_0$  
between modes excited in consecutive orders of perturbation theory.
The delocalization for $k_0 \rightarrow 0$ however is
due to an approaching of resonances $n\omega_{k_0} \rightarrow \omega_{nk_0}$ for some
integer $n\ge 2$, which are the Ford resonances \calleq{acres}.

Most importantly, again for each value of $\alpha$ and $\epsilon$ we can identify a critical
seed wave number 
\begin{equation}
\label{alphakc1}
{k}_{c1} = \frac{\alpha^2 \epsilon}{2 \pi^3}\;. 
\end{equation}
For $k_0 > {k}_{c1}$
$q$-breathers are localized, while for $k_0 < {k}_{c1}$ they delocalize. This delocalization
threshold (\ref{thresholdbeta})  is expected to translate into a delocalization threshold
of a corresponding FPU trajectory.
If instead we use the estimate (\ref{alphadelocstrong}),
we arrive at a critical seed wave number 
\begin{equation}
\lab{alphakc2}
{k}_{c2} = \alpha^{1/2} \epsilon^{1/4} \;.
\end{equation}

\subsection{Some consequencies}

Let us discuss the connection between the time scales $\tau_1$ and $\tau_2$, and
the localization-delocalization threshold for $q$-breathers. 
According to our understanding, for  $k_0 > {k}_{c1}$ it follows 
$\tau_2 \gg \tau_1$. For $k_0 \approx {k}_{c1}$
we reach $\tau_2 \approx \tau_1$, and when further decreasing $k_0 < {k}_{c1}$, both scales stay close, but
diverge in the limit of long wavelength. Note that the critical number $k_{c1}$ has to be replaced
with $k_{c2}$ if one takes the scaling results for strongly localized $q$-breathers.
These conclusions are only reasonable if we start a trajectory close to 
the $q$-breather. Thus, if the $q$-breather is almost (or completely) delocalized, an
initial condition similar to the one considered by FPU (which is strongly localized
in $q$-space) is at large distance from the corresponding $q$-breather, and the
relaxation properties of the FPu trajectory may become very different.

For a large (possibly infinite) system, at some given energy density and nonlinearity parameter,
a $q$-breather necessarily delocalizes for some small but finite seed wave number $k_0$,
signalling the breakdown of the corresponding dressed phonon picture familiar from solid state physics.
That may seem strange, since it is textbook wisdom that acoustic phonons have lifetimes
which increase with increasing wavelength. Yet the two results perfectly match.
Indeed, the lifetime measures the relaxation time of a bare phonon (normal mode) injected into
the system, in units of its period. These lifetimes indeed grow with increasing wavelength, since
the coupling to shorter wavelength modes is inversely proportional to the third ($\alpha$)
or fourth ($\beta$) power of the wavelength. At the same time the couplings become more and more
resonant, so that the new state which is reached after that waiting time becomes more and more
different from the bare phonon state (normal mode).

Finally, we emphasize that quantum phonon physics usually considers the decay of one phonon
into two phonons of longer wavelength. However in the classical limit, considered here,
the occupation number of the initial phonon state is large compared to one, and the leading
order decay is given by annihilating two ($\alpha$) or three ($\beta$) phonons and creating
one with corresponding shorter wavelength, in full agreement with the numerical simulations
of FPU, and with all the other studies of the corresponding classical problem.

\section{The normal form approach to the energy cascade}

The picture emerging from the above mentioned approach to the FPU problem via $q$-breathers leads to 
the conclusion that the perturbative continuation of a normal mode of wavenumber $k_0$ of the linearized system \calleq{H1} 
converges to an exact periodic solution of the full nonlinear system when $k_0$ is larger than
some boundary or critical value $k_{c}$, 
the latter quantity being a function of the specific energy $\epsilon$ of the system. 
Under the same condition, $k_0>k_c$, 
the modal energy spectrum of such a periodic orbit is exponentially localized in $k$-space. Thus, in the corresponding initial value problem,
when a mode of wavenumber $k_0>k_c$ 
is initially excited, the FPU paradox can be explained in terms of closeness of the 
initial datum to an exact periodic orbit of the system.
On the other hand, the continued orbit of a mode with wavenumber $k_0$ smaller or much 
smaller than $k_c$, when it exists, is expected 
to be delocalized in $k$-space, and the dynamics of the 
corresponding initial value problem,
with initial excitation of a mode with wavenumber $k_0\ll k_c$, becomes different.
In the latter case the energy initially given to the mode with $k_0\ll k_c$ is 
expected to be transferred to higher wavenumber modes within some characteristic time $\tau_c$.

\begin{figure}
{
\resizebox*{0.7\columnwidth}{!}{\includegraphics{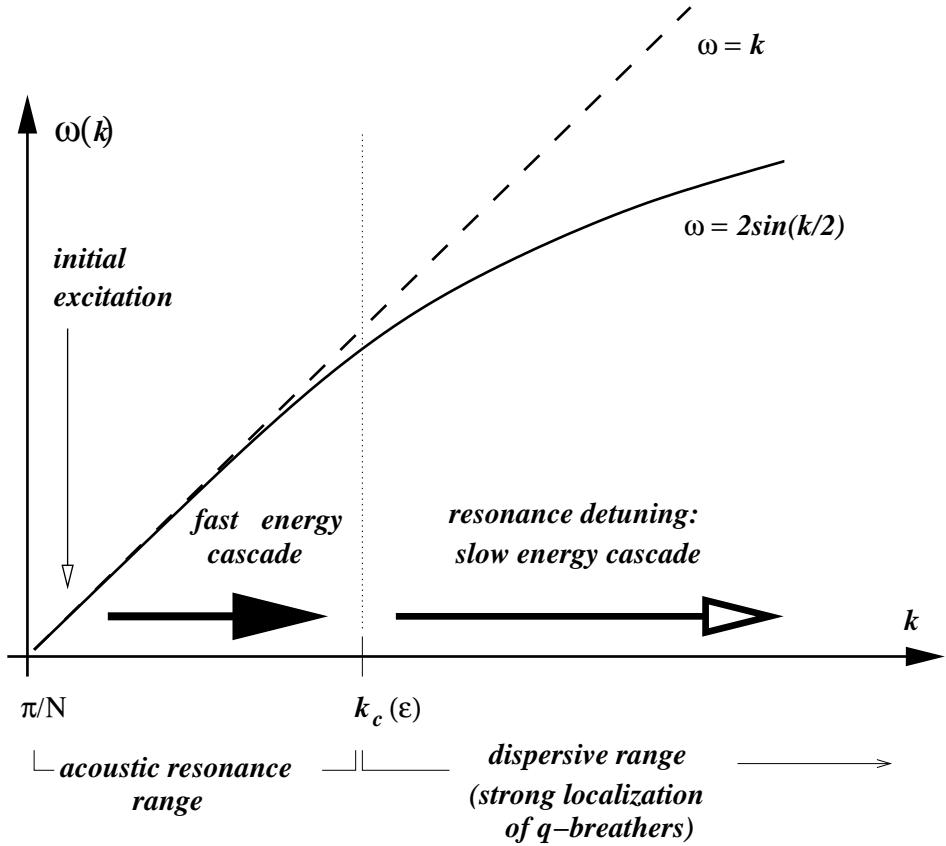}}
}
\caption{A synoptic diagram describing the energy cascade in FPU (the thick solid curve 
is the graph of the dispersion relation
\calleq{disprel}). 
The vertical thin-dotted line marks the position of the critical wavenumber $k_c$.
The long-dashed straight line is the graph of the dispersion relation linearized around $k=0$, which
yields the Ford resonances \calleq{acres}. 
The fast energy cascade takes place on the time scale $\tau_c(\epsilon)$, whereas 
a second slower 
stage of the cascade is expected to continue into the
more distant part of the wavenumber space, as observed numerically.}
\label{Diag}
\end{figure}

The qualitative outcome is sketched in Fig.\ref{Diag}.
There exists a wavenumber interval in $k$-space, $\pi/N\lesssim k\lesssim k_c(\epsilon)$, 
that one can refer to as the \emph{acoustic resonance range}.
The energy initially injected inside such a wavenumber range cascades up to the border $k\approx k_c$, 
within a characteristic time $\tau_c$,
at least if the specific energy $\epsilon$ is small enough. The subsequent transfer of energy inside the range 
$k_c(\epsilon)\lesssim k\lesssim \pi$,
that one can refer to as the \emph{dispersive range},
is slowed down by the detuning from the almost complete acoustic resonance
of the corresponding unperturbed frequencies. 
The path to energy equipartition is expected to be 
completed on some time scale(s) much longer than $\tau_c$. 
Such a picture emerged for the first time in \cite{Sh,BGG,BKL}, where estimates of $k_c(\epsilon)$ and of 
$\tau_c(\epsilon)$, were provided,
mainly for the $\alpha$-model. Moreover, it 
was numerically shown there that
the modal energy spectrum of the system displays a \emph{Wien-like} exponentially decreasing tail inside the dispersive 
range, namely $E(k)\sim e^{-k/k_c}$ for $k>k_c$,
as originally observed in \cite{GS} and stressed again in \cite{Fetal}.

The fundamental scaling laws
of the problem (i.e. the dependence of $k_c$ and $\tau_c$ on $\epsilon$) can be obtained through the 
approach based on the quasi-resonant normal form
construction introduced in \cite{Sh} and developed in \cite{P2,PB1,PB2,BP1,BP2}. 
The idea starts from the expansion \calleq{omexp} of the dispersion relation \calleq{disprel}, 
namely
$\omega_q=2\sin(k_q/2)=k_q-k_q^3/24+\dots$ (recall here that $k_q=\pi q/(N+1)$, where $q=1,2\dots$), 
which holds if $k_q$ is small.
The quadratic part $h_2$ of the Hamiltonian \calleq{H3} can thus be rewritten as
$h_2=\sum_{q=1}^N\omega_q|z_q|^2=\sum_{q=1}^Nk_q|z_q|^2+\sum_{q=1}^N\Omega_q|z_q|^2$,
with $\Omega_q\equiv \omega_q-k_q$.
For initial excitations of acoustic modes, the sum $E_h=\sum_{q=1}^Nk_q|z_q|^2$ can be regarded
as the unperturbed Hamiltonian of the problem. Such a splitting holds if low modes only are involved in the dynamics and 
consistency is recovered at the end, showing that the fraction of modes $k_c$ sharing the energy turns out to be small when the specific 
energy $\epsilon$ is small. One now introduces a time-dependent canonical change of variables $z\mapsto\zeta,$ the new variables $\zeta_q(t)$ being defined by 
$z_q=\zeta_q\exp(-ik_qt)$ ($q=1,\dots,N$), in terms of which the Hamiltonian of the problem is mapped to
a new Hamiltonian $K=\overline{K}(\zeta,\zeta^*)+R(\zeta,\zeta^*,t)$, where
\begin{eqnarray}
 \overline{K}(\zeta,\zeta^*)& = & \sum_{q=1}^N\Omega_q|\zeta_q|^2+\frac{\alpha}{4\sqrt{N+1}}\sum_{j,q,l=1}^N\delta_{j+q,l}
 \sqrt{\omega_j\omega_q\omega_l}(\zeta_j^*\zeta_q^*\zeta_l+c.c.)+ \nonumber\\
 && +\frac{\beta}{8(N+1)}\sum_{j,q,l,m=1}^N\delta_{j+q+l,m}\sqrt{\omega_j\omega_q\omega_l\omega_m}\ 
 (\zeta_j^*\zeta_q^*\zeta_l^*\zeta_m+c.c.)+\nonumber\\
 && +\frac{3\beta}{32(N+1)}\sum_{j,q,l,m=1}^N\delta_{j+q,l+m}\sqrt{\omega_j\omega_q\omega_l\omega_m}\ 
 (\zeta_j^*\zeta_q^*\zeta_l\zeta_m+c.c.)
\lab{H4}
\end{eqnarray}
and the remainder $R$ collects all the terms that explicitly depend on time through
some oscillating exponential. The time--average $\overline{R}$ of $R$, the $\zeta_k$'s being held constant, is zero, and
one can study, in a first approximation, the dynamics associated to the Hamiltonian \calleq{H4}
which is precisely the \emph{quasi-resonant normal form Hamiltonian of the FPU problem}.
The rigorous justification of such an approximation, i.e. the neglecting of the remainder $R$,
is a difficult task, and precise statements are found in \cite{BP1,BP2}. 
Before passing to a direct inspection of the equations of motion associated to the Hamiltonian \calleq{H4}
for the $\alpha$--model and for the $\beta$--model, we observe that, as a consequence of averaging,
the unperturbed Hamiltonian $E_h=\sum_{q=1}^N k_q|\zeta_q|^2$ commutes with the normal form Hamiltonian \calleq{H4}.
The second integral $E_h$ approximately coincides with the sum of the harmonic energies of the modes, 
at least for excitations involving acoustic (low frequency) modes only. 

Let us now consider the $\alpha$ and $\beta$ models
separately. In both cases, one
\emph{supposes} that only a fraction of the order $k_c\ll 1$ of acoustic modes is involved in the dynamics, so that $\zeta_q\simeq0$ for $q>Nk_c$ ($k_c$ 
to be estimated below). One then introduces the new \emph{noncanonical} variables
\begin{equation}
 u_q=\sqrt{k_q}\ \zeta_q\ ,
\lab{uk}
\end{equation}
in terms of which the second integral $E_h$, i.e. the sum of the acoustic harmonic energies, becomes 
\begin{equation}
E_h=\sum_{k=1}^N|u_k|^2\equiv N\varepsilon\ .
\lab{Eh}
\end{equation}
In the latter formula we have defined the harmonic specific energy $\varepsilon$ as the acoustic harmonic energy per degree of freedom.
Of course, as far as the perturbative approach holds, one has $\varepsilon\simeq\epsilon$.
Notice that the quantity $|u_q|^2$ is the approximate harmonic energy of the mode $q$.

\subsection{$\alpha$--model}
 
The equations of motion of the $\alpha$-model, in terms of the $u$ variables \calleq{uk} take the form
\begin{equation}
 i\dot{u}_q=-\frac{k_q^3}{24}\ u_q+\frac{k_q\alpha}{4\sqrt{N+1}}
 \left[\sum_{l=1}^{q-1}u_{q-l}u_l
 +2\sum_{l=1}^{N-q}u^*_{l}u_{q+l}\right]\ .
\lab{aeq2}
\end{equation} 
The fundamental step consists now in recognizing that, through the scaling
transformation $(u_q,k_q,t)\mapsto(v_q,x_q,T)$, defined by
\begin{equation}
 u_q=\varepsilon^{1/2}v_q\ \ ,\ \ k_q=\alpha^{1/2}\varepsilon^{1/4}x_q\ \ ,\ \ 
 t=\alpha^{-3/2}\varepsilon^{-3/4}T\ \ ,
\lab{ascal}
\end{equation}
the equations of motion \calleq{aeq2} and the second integral \calleq{Eh} transform to
\begin{eqnarray}
 && i\frac{dv_q}{dT}=-\frac{x_q^3}{24}\ v_q+
 \frac{x_q}{4\sqrt{N+1}}
 \left[\sum_{l=1}^{q-1}v_{q-l}v_l
 +2\sum_{l=1}^{N-q}v^*_{l}v_{q+l}\right]\ \ ,\lab{aeq3} \\
 && \frac{1}{N}\sum_{q=1}^N|v_q|^2=1\lab{Eh2}\ \ .
\end{eqnarray}
The latter system no longer depends parametrically on $\alpha$ and/or $\varepsilon$, the only parameter
entering the equations of motion being $N$. We now define an effective Reynolds number $\mathcal{R}(x_q)$ 
as the ratio of the nonlinear to the linear term appearing on the right hand side of \calleq{aeq3}, both terms being squared and averaged over the phases
of the active modes (see \cite{P2}). Thus $\mathcal{R}(x_q)$ measures, for any mode $q$, the relevance of mode coupling (nonlinearity) with respect to
dispersion. It follows that,  
\emph{under the hypothesis that the phases of the active modes are randomly distributed}, $\mathcal{R}(x)$ decreases below one at $x^c\sim 1$, independent of $N$.
From the equations of motion \calleq{aeq3} it follows that the time scale $T_c$ associated with
the dynamics of the mode with rescaled wavenumber $x^c$ is given by the reciprocal of its frequency, namely 
$T_c\approx (x^c)^{-3}$. Thus also $T_c\sim1$ does not depend on $N$ and, from the second and third of the definitions \calleq{ascal} 
the estimates
\begin{eqnarray}
 && k_c \sim \sqrt{\alpha}\ \varepsilon^{1/4}\ ,\lab{kcalpha}\\
 && \tau_c\sim\alpha^{-3/2}\varepsilon^{-3/4}\ ,\lab{tcalpha}
\end{eqnarray}
follow. Notice that the scaling of $k_c$ coincides with that of $k_{c2}$, equation \calleq{alphakc2}. 

\subsection{$\beta$--model}

For what concerns the $\beta$--model we will write down the main formulas, since the reasonings are much the same 
as those made for the $\alpha$--model. The equations of motion of the system written in terms of the $u$ variables \calleq{uk} are
\begin{eqnarray}
 && i\dot{u}_q=-\frac{k_q^3}{24}\ u_q+\frac{k_q\beta}{8(N+1)}\left[
 \sum_{2\leq l+m\leq q-1}u_lu_mu_{q-l-m}+\right.\nonumber\\
 && \left.+3\sum_{2\leq l+m\leq N-q}u_l^*u_m^*u_{l+m+q}+
 3\sum_{q+1\leq l+m\leq N+q}u_{l+m-q}^*u_lu_m\right]\ \ ,
 \lab{beq2}
\end{eqnarray}
while the conservation law \calleq{Eh} is exactly the same. The scaling transformation 
$(u_k,\xi_k,t)\mapsto(v_k,x_k,T)$ is now given by
\begin{equation}
 u_q=\varepsilon^{1/2}\ v_q\ \ ,\ \ \xi_q=(\beta\varepsilon)^{1/2}\ x_q\ \ ,\ \ 
 t=(\beta\varepsilon)^{-3/2}T\ \ ,
\lab{bscal}
\end{equation}
through which the equations \calleq{beq2} transform to
\begin{eqnarray}
 && i\frac{dv_q}{dT}=-\frac{k_q^3}{24}\ v_q+\frac{k_q}{8(N+1)}\left[
 \sum_{2\leq l+m\leq q-1}v_lv_mv_{q-l-m}+\right.\nonumber\\
 && \left.+3\sum_{2\leq l+m\leq N-q}v_l^*v_m^*v_{l+m+q}+
 3\sum_{q+1\leq l+m\leq N+q}v_{l+m-q}^*v_lv_m\right]\ \ ,
 \lab{beq3}
\end{eqnarray}
togheter with the conservation law $\sum_{q=1}^N|v_q|^2/N=1$. The ratio nonlinearity/dispersion $\mathcal{R}(x_q)$ is then defined in a 
way completely analogous to what done above
for the $\alpha$--model and, under the hypothesis that the phases of the active modes are randomly distributed one gets
that $\mathcal{R}(x)\lesssim 1$ as $x\gtrsim 1$. Thus, also for the $\beta$-model $x^c\sim 1$, $T_c\sim 1/(x^c)^3\sim 1$,
both independent of $N$, and from the scaling transformations \calleq{bscal} one gets the estimates
\begin{eqnarray}
 && k_c \sim \sqrt{\beta\varepsilon}\ ,\lab{kcbeta}\\
 && \tau_c\sim(\beta\varepsilon)^{-3/2}\ .\lab{tcbeta}
\end{eqnarray}
Here again we stress that $k_c$ displays the same scaling of $k_{c2}$ given in \calleq{betakc2}.

\subsection{Comments}

First of all, 
note that the obtained scaling properties of $k_c$ correspond to the scaling properties 
of $k_{c2}$ for $q$-breathers, for the $\alpha$ and $\beta$ models, respectively. 
This is quite natural: the region where $q$-breather miss their \emph{strong} localization property
is the one where the acoustic resonance effectively works, leading to a partial relaxation 
of the system.
Recall however that $k_{c2}$ is obtained using the scaling relationships for strongly localized
$q$-breathers, while a more careful analysis yields estimate $k_{c1}$ which scale
differently and are lower than $k_{c2}$. That leaves room for further investigations and comparisons.

As a further comment, making reference to Fig.\ref{Diag}, we stress that in order to observe the scenario depicted above
(fast cascade up to $k_c$ and slow one from $k_c$ on) it is necessary that $\pi/N\ll k_c(\epsilon)\ll\pi$,
so that the acoustic resonance range has a small but finite size, and the dispersive range occupies the most part of $k$-space.
The upper bound on $k_c$ gives a threshold value of the specific energy above which a fast trend to equipartition occurs.
The lower bound on $k_c$ yields instead an estimate of the minimum number $N$ of degrees of freedom necessary
to \emph{open} an acoustic range in the system at a fixed total energy $E$. For the $\alpha$-model one gets
$N\gg \alpha^{-2/3}E^{-1/3}$, whereas for the $\beta$-model it must be $N\gg 1/(\beta E)$ (neglecting numerical factors).

We want to stress here that the hypothesis of random phases seems to play a crucial role in the above derivation,
in particular for what concerns the dependence of $k_c$ and $\tau_c$ on the number $N$ of degrees of freedom of the system.
We anticipate that when one replaces such an hypothesis with that of coherent phases, for example,
the estimates \calleq{kcalpha}--\calleq{tcbeta} hold the same but for the replacement of the specific energy
$\varepsilon$ with the harmonic energy $E_h=N\varepsilon$, which approximately coincides with the total energy of the system. 
Whether the phases of the active modes are randomly distributed
or not seems to depend strongly on what one chooses initially: if one initially excites a small fraction of acoustic modes,
randomness or coherence in the phases of the modes initially excited
is recalled by the system, and such a strong memory effect is another signal of quasi-integrable behaviour \cite{BLP}.

It must be stressed that the scaling \calleq{kcalpha}, first obtained by Shepelyansky, 
both numerically and analytically, has been confirmed numerically in two successive works
by Berchialla et al. \cite{BGG} and Biello et al. \cite{BKL}. Notice that the characteristic time $\tau_c$
\calleq{tcalpha} was first written down in \cite{Sh} on the basis of dimensional arguments and 
interpreted as the largest Lyapunov exponent of the system, whereas it was first measured in 
\cite{BKL} and identified
as the time scale at which energy is transferred to short wavelengths.
Moreover, the analytical explanation of the exponential tail in the modal energy spectrum has been given in
\cite{P1,P2,PB1,PB2,BP1,BP2}, where it has been shown that the FPU quasi-resonant normal form dynamics in real space
is described by two KdV equations for the $\alpha$-model and by two modified KdV equations for the $\beta$-model.
Peculiar integrability properties of the KdV equations yield then an exponentially 
decreasing modal energy spectrum; see in particular \cite{P1} where, based on soliton theory, a formula 
describing all the spectrum of the $\alpha$-model (not only the high frequency tail) is given.

Another consequence of the analysis carried out in terms of normal forms is the following.
At least in the case of the $\alpha$ and $\beta$-models, the normal form being integrable,
it turns out that the first stage of the dynamics, namely up to times of the order of $\tau_c$,
is almost \emph{regular}. Thus $\tau_c$ has to be interpreted just as a redistribution time,
and not as a stochastization time scale. As a good example one can think of the Toda model,
whose dynamics, for low specific energies, displays all the features of the $\alpha$-model.
However, such a similarity breaks down on some time scale, after which the dynamics of the Toda chain 
goes on to be regular (quasi periodic with zero Lyapunov exponents) while that of the $\alpha$-model
develops chaotic features (local stochasticity and a nonzero Lyapunov spectrum); see \cite{GPP}.
The onset of stochasticity in the FPU dynamics takes place on time scales 
surely longer than $\tau_c(\epsilon)$. 

\section{Resonance overlap criteria}

The idea of the resonance overlap is to obtain parameter estimates where
a certain resonance in the nonlinear system may develop, leading to a 
strong change in the dynamics of the chain, presumably to a fast equipartition.
The main ideas are published in \cite{IC} and \cite{C1}, and we are going to present
here a very simple and qualitative way to derive the final results for the $\beta$-FPU case.

The selfinteraction of an initially excited mode with seed wave number $k_0$ is described
by
\begin{equation}
\ddot{Q}_{k_0} + \omega^2_{k_0} \sim \frac{\beta}{N} \omega^4_{k_0} Q^3_{k_0}
\;.
\label{selfinteraction}
\end{equation}
This is an anharmonic oscillator problem, where the frequency depends on
the energy \cite{AHN}.
Consequently the frequency shift of this anharmonic oscillator problem, when excited
to a given energy $E=N \epsilon $, is given by
\begin{equation}
\delta \omega \sim \beta \omega_{k_0} \epsilon
\;.
\label{ICshift}
\end{equation}
When this shifted frequency matches some other frequency, a resonance may develop, leading
to a resonant decay of the originally excited mode. 
The entire final result will now depend on the choice of the frequency to be compared with,
which depends on the model.
Izrailev and Chirikov chose the
seemingly natural frequency given by the separation of the normal mode frequencies due to
the finite size of the system. Such a frequency separation is $\Delta_{IC} \sim 1/N$ for
long wavelength modes. Comparing it with the nonlinear frequency shift $\delta \omega$ one
arrives at the threshold condition
\begin{equation}
\beta \omega_{k_0} \epsilon \sim \frac{1}{N}
\;.
\label{ICcriterion}
\end{equation}
Clearly this condition is not entirely depending on intensive quantities. 
Furthermore no correspondence to the above derived delocalization thresholds for $q$-breathers and
scaling relations for normal forms can be found. Finally note, that for the longest
wavelength mode $q=1$ it follows that the critical energy density according to IC is independent
of the system size: $\epsilon \sim 1/\beta$. However $q$-breathers and normal forms
predict $\epsilon \sim 1/(\beta N^2)$ in that case. Thus delocalization happens for much smaller
energy densities than predicted by the IC criterion.

The reason for the above discrepancy is the chosen resonance.
Indeed, as we know from the perturbation theory of $q$-breathers, and from the analysis
of normal forms, the leading order resonances are those between the third harmonic
of the seed normal mode and the normal mode with three times smaller wavelength:
$\Delta_{Sh} = \omega_{3k_0} - 3\omega_{k_0}$. Within the context
of resonance overlap this was first recognized by Shepelyansky \cite{Sh}.
For long wavelength it follows from \calleq{omexp}, that we may approximate this frequency by
$\Delta_{Sh} \sim k_0^3$. Comparing 
that with $\delta \omega$ and using $\omega_{k_0} \approx k_0$
we find the correct threshold condition
\begin{equation}
\frac{\beta \epsilon}{k_0^2} \sim 1
\;.
\label{Shcriterion}
\end{equation}
This condition is in excellent agreement with the scaling properties of strongly localized
$q$-breathers, and with the scaling properties of resonant normal forms.
We note, that the slightly more cumbersome analysis of the $\alpha$-FPU case
yields exactly the same conclusions (compare \cite{Sh} and \cite{FIKa}). 

\section{Conclusions}

We find that all three approaches -
resonant overlap, periodic orbits (for strong localization) 
and resonant normal forms - yield identical scaling relations,
if the proper resonance
condition is identified and exploited.
These scaling relations are depending on intensive quantities only. 
Consequently the
above formulated FPU problems can be explained in the following way.
There are two time scales, since the excitation of a normal mode below
the delocalization threshold is close to an exact periodic orbit - a $q$-breather
- which localizes exponentially in normal mode space.
It will need a short time $\tau_1$ to approach the $q$-breather.
The excited trajectory will then spend a long time $\tau_2$ in a neighbourhood of the periodic orbit.
When approaching the delocalization threshold, the two time scales merge,
and equipartition is established on a single time scale. The thresholds
depend on intensive quantities only, and thus the FPU effect is robust
when considering the limit of infinitely large systems - but one has to remember
to use intensive parameters. E.g. the often used scheme of exciting 
the mode with $q_0=1$ but increasing the system size would lead to a
change of the intensive quantity of a wave number $k_0$ or the frequency
$\omega_{k_0}$. 
An open question concerns whether the different scaling $k_{c1}$ as compared to $k_{c2}$
obtained for $q$-breathers, is also observed within the normal form approach.
Another open point is the analytical estimate of the times $\tau_1$ and $\tau_2$
and their relation to $\tau_c$. 

Finally the generalization to higher spatial dimensions does not seem
to bring about qualitatively new features, at least from the point of view of $q$-breathers.
However, we think that
much more detailed studies have to be conducted to be on the more
safe side with such a statement (see \cite{Be} for 
some recent studies). 

To end with,
we point out that 
Lichtenberg \cite{Li} performed recently an estimate of the (weak) stochasticity
threshold for the $\alpha$-model, similar to previous studies for the $\beta$-model
\cite{LLL}. Surprisingly the results for the $\alpha$-model are in full agreement
with the scaling laws following from the analysis of $q$-breathers and normal forms.

\acknowledgments
A. P. thanks G. Benettin, L. Galgani and A. I. Neishtadt for useful comments and discussions on the subject.


\begin{thebibliography}{0}


\bibitem{FPU} E. Fermi, J. Pasta and S. Ulam, in
\emph{Enrico Fermi -- Note e Memorie (collected papers)}, vol. II (USA 1939--1954), Accademia Nazionale dei Lincei, Roma (Italy), 
and The University of Chicago Press, Chicago (USA) 1965, p. 978. 
 
\bibitem{Ford92}
J. Ford, Phys. Rep. {\bf 213} 271-310 (1992). 

\bibitem{GS} L. Galgani and A. Scotti, Phys. Rev. Lett. {\bf 28}, 1173--1176 (1972).

\bibitem{F} J. Ford, J. Math. Phys. {\bf 2} 387--393 (1961).

\bibitem{IC} F. M. Izrailev and B. V. Chirikov, Sov. Phys. Dokl. {\bf 11}, 30-32 (1966).

\bibitem{Sh} D. L. Shepelyansky, Nonlinearity {\bf 10}, 1331--1338 (1997).

\bibitem{FIKa} S. Flach, M. V. Ivanchenko, and O. I. Kanakov, Phys. Rev. Lett. {\bf 95}, 064102/1-4 (2005).

\bibitem{FIKb} S. Flach, M. V. Ivanchenko, and O. I. Kanakov, Phys. Rev. E {\bf 73}, 036618/1-14 (2006). 

\bibitem{IKMF} M. V. Ivanchenko, O. I. Kanakov, K. G. Mishagin and S. Flach,
Phys. Rev. Lett. {\bf 97}, 025505/1-4 (2006).

\bibitem{KFIM}  O. I. Kanakov, S. Flach, M. V. Ivanchenko and K. G. Mishagin, 
Phys. Lett. A {\bf 365}, 416-420 (2007).

\bibitem{PF} T. Penati and S. Flach, Chaos {\bf 17}, 023102/1-16 (2007).

\bibitem{FKMI} S. Flach, O. I. Kanakov, K. G. Mishagin and M. V. Ivanchenko,
Int. J. Mod. Phys. B, in press.

\bibitem{P1} A. Ponno, Europhys. Lett. {\bf 64} 606--612 (2003).

\bibitem{P2} A. Ponno, in the Proceedings of the Carg\'ese Summers School 2003 on 
\emph{Chaotic Dynamics and Transport in Classical and Quantum Systems}, edited by P. Collet et al., 
Kluwer Achademic Publishers, Netherlands, 2005, p. 431.

\bibitem{PB1} A. Ponno and D. Bambusi, in the Proceedings of the International Conference on \emph{Symmetry and Perturbation Theory   2004}, 
edited by G. Gaeta et al., World Scientific Publishing, Singapore, 2005, p. 263.

\bibitem{PB2} A. Ponno and D. Bambusi, Chaos {\bf 15} 015107 (2005).

\bibitem{BP1} D. Bambusi and A. Ponno, Comm. Math. Phys. {\bf 264}, 539--561 (2006). 

\bibitem{BP2} D. Bambusi and A. Ponno, in the
Review Articles on Fermi-Pasta-Ulam Experiment grown out of the meeting \emph{FPU fifty years later} held in Rome (Italy) on May, 
7th 2004 (to appear on a Springer Book edited by G. Gallavotti);
available on the web site http://ipparco.roma1.infn.it/index.html.

\bibitem{ZK} N. J. Zabusky and M. D. Kruskal, Phys. Rev. Lett. {\bf 15} 240--243 (1965).

\bibitem{BGP} L. Berchialla, A. Giorgilli and S. Paleari, Phys. Lett. A {\bf 321}, 167--172 (2004).

\bibitem{Fetal} F. Fucito \emph{et al.}, J. Physique {\bf 43}, 707-713 (1982).

\bibitem{ford63} J. Ford and J. Waters, J. Math. Phys. {\bf 4}, 1293-1306 (1963).

\bibitem{Ly} A. Lyapunov, \emph{Probl\`eme g\'en\'eral de la stabilit\'e du mouvement}, 
published by the Mathematical Society of Kharkov, 1892 (Russian); available in
\'Editions Jacques Gabay, Paris (France) 1988 (French), reprint of the French translation from Russian appeared on the Annales de 
la Facult\'e des Sciences de Toulouse 2\`eme s\'erie, tome IX, 1907.

\bibitem{LLL} J. De Luca, A. Lichtenberg and M. A. Lieberman, Chaos {\bf 5}, 283--297 (1995).

\bibitem{BGG} L. Berchialla, L. Galgani, and A. Giorgilli, DCDS-A {\bf 11}, 855-866 (2004).

\bibitem{BKL} J. A. Biello, P. R. Kramer, Y. V. Lvov, Stages of energy transfer in the FPU model,
DCDS Supplements, 2003; special number devoted to the
Proceedings of the Fourth International Conference on Dynamical Systems and Differential Equations,
May 24--27, 2002, Wilmington, NC, USA, p. 113.

\bibitem{BLP} G. Benettin, R. Livi and A. Ponno, work in progress.

\bibitem{GPP} A. Giorgilli, S. Paleari and T. Penati, DCDS-B {\bf 5}, 991-1004 (2005).

\bibitem{C1} B. V. Chirikov, Internal Report n. 267 of the Nuclear Institute of the Siberian Section of the USSR Academy of Sciences, 
Novosibirsk, 1969; CERN Translation n. 70--41, Geneva 1971.

\bibitem{AHN} A. H. Nayfeh, Introduction to Perturbation Techniques (Wiley, New York, 1993).

\bibitem{Be} G. Benettin, Chaos {\bf 15}, 015108 (2005).

\bibitem{Li} A. Lichtenberg, private communication.

\end{thebibliography}
\end{document}